\documentclass[preprint,showpacs,aps]{revtex4}
\newcommand{\UQ}{ARC Centre of Excellence for Quantum-Atom Optics, 
School of Physical Sciences, University of Queensland, Brisbane, 
QLD 4072, Australia.}

\usepackage{graphicx,psfrag}
\usepackage{bm}
\newcommand{\etal}{{\em et al.}}
\newcommand{\e}{\mbox{e}}
\newcommand{\pr}{Phys. Rev. }
\newcommand{\jpb}{J. Phys. B }
\newcommand{\opex}{Opt. Express }
\newcommand{\jpa}{J. Phys. A }

\newcommand{\qso}{Quantum Semiclass. Opt. }
\newcommand{\job}{J. Opt. B }

\begin{document}
\title{Bright bichromatic entanglement and quantum dynamics of sum frequency generation}

\author{M.~K. Olsen and A.~S. Bradley}
\affiliation{\UQ}
\date{\today}

\begin{abstract}

We investigate the quantum properties of the well-known process of sum frequency generation, showing that it is potentially a very useful source of non-classical states of the electromagnetic field, some of which are not possible with the more common techniques. We show that it can produce quadrature squeezed light, bright bichromatic entangled states and symmetric and asymmetric demonstrations of the Einstein-Podolsky-Rosen paradox. We also show that the semiclassical equations totally fail to describe the mean-field dynamics when the cavity is strongly pumped.  

\end{abstract}


\maketitle

\section{Introduction}
\label{sec:intro}

The nonlinear optical process of sum frequency generation (SFG), also known as non-degenerate upconversion or frequency summation~\cite{Boyd}, has been known and investigated for a number of years, going back at least to theoretical investigations in the seminal paper of Armstrong \etal~\cite{Armstrong} and experimental realisation by Bass \etal~\cite{Bass}. The process has many uses outside the quantum optics community, such as, to name only a few, surface vibrational spectroscopy of molecules~\cite{surface}, two-dimensional vibrational spectroscopy~\cite{Cho}, studies of liquid interfaces~\cite{liquid1,liquid2}, low noise optical tomography~\cite{tomography}, and the investigation of powder supported catalysts~\cite{catalyst}.

However, despite the wide number of what we may term classical uses for this process, very little attention seems to have been paid to its quantum properties, especially in terms of entanglement and quadrature squeezing. For the intracavity process, a notable exception is the theoretical work of Eschmann and Reid~\cite{Eschmann}, who predicted sub-Poissonian photon statistics in the high frequency mode and in the sum of the two low frequency modes. More attention has been paid to type II second harmonic generation, where the non-degeneracy is in polarisation rather than in frequency, for example work by Jack, Collett and Walls~\cite{MWJ}, and by Andersen and Buchhave~\cite{Ulrik}. Four-wave mixing is also a process which can lead to sub-Poissonian fluctuations in intensity sums~\cite{gcb}, as well as a source of entangled beams of either photons~\cite{FWMphotons} or atoms~\cite{mjd}. This leads us to expect that sum frequency generation may also be a source of entangled output beams. 

In this work we will first introduce the full Hamiltonian of the intracavity process before examining the properties of the interaction Hamiltonian considered in isolation. We will then analyse the outputs of the full intracavity system in terms of squeezing and entanglement. We will show that SFG is in fact a versatile source of entanglement resources which can be easily tuned to entangle beams which have large frequency differences and produce entanglement between output beams at different intensities. The tunability of the process also allows us to predict that the same device could be used to demonstrate both symmetric and asymmetric steering~\cite{steering}, which may have applications in the field of quantum cryptography. We will show that this process is potentially very useful for the field of continuous variable quantum information.

\section{System and Hamiltonian}
\label{sec:Ham}

The basic interaction is that of a photon at $\omega_{1}$ combining with a photon at $\omega_{2}$ to produce a photon at $\omega_{3}\:(=\omega_{1}+\omega_{2})$, mediated by a second order, $\chi^{(2)}$, nonlinearity. (For an accessible description of advances in the use of $\chi^{(2)}$ materials, see Hanna~\cite{Hanna}.) 
The full Hamiltonian describing this interaction inside a triply resonant optical cavity and the interaction of the cavity fields with the external fields may be written as
\begin{equation}
{\cal H} = {\cal H}_{int}+{\cal H}_{pump}+{\cal H}_{res},
\label{eq:Hfull}
\end{equation}
where the interaction Hamiltonian in the appropriate rotating frame is
\begin{equation}
{\cal H}_{int} = i\hbar\kappa\left[\hat{a}_{1}^{\dag}\hat{a}_{2}^{\dag}\hat{a}_{3}-\hat{a}_{1}\hat{a}_{2}\hat{a}_{3}^{\dag}\right],
\label{eq:Hint}
\end{equation}
the pumping Hamiltonian is
\begin{equation}
{\cal H}_{pump} = i\hbar\sum_{i=1}^{2}\left[\epsilon_{i}\hat{a}_{i}^{\dag}-\epsilon_{i}^{\ast}\hat{a}_{i}\right],
\label{eq:Hpump}
\end{equation}
and the reservoir damping Hamiltonian is
\begin{equation}
{\cal H}_{res} = \hbar\sum_{i=1}^{3}\left[\hat{\Gamma}_{i}\hat{a}_{i}^{\dag}+\hat{\Gamma}_{i}^{\dag}\hat{a}_{i}\right].
\label{eq:Hres}
\end{equation}
In the above, $\hat{a}_{i}$ is the bosonic annihilation operator for the mode at frequency $\omega_{i}$, $\kappa$ represents the effective second order nonlinearity, the $\epsilon_{i}$ are the classical pumping laser amplitudes at the respective frequencies, and the $\hat{\Gamma}_{i}$ are the annihilation operators for bath quanta, representing losses through the cavity mirror.

\section{Equations of motion and Hamiltonian dynamics}
\label{sec:Positivepee}

We will begin by giving the equations of motion which result from considering the interaction Hamiltonian in isolation and then add cavity loss and pump terms below, in section~\ref{sec:cavidade}, as the addition of these is a trivial matter. We note here that the approach we use is not expected to give completely accurate predictions for the system operating in the travelling wave configuration as it does not account for such physical features as dispersion in the nonlinear medium. However, it does give an approximate idea and is useful for understanding the quantum dynamics which the Hamiltonian makes possible. 
Following the usual procedures~\cite{Danbook,GardinerQN}, we may map the interaction Hamiltonian onto a Fokker-Planck equation for the P-representation pseudoprobability distribution of the system~\cite{Glauber,Sudarshan}. Making the operator correspondences $\hat{a}_{i}\leftrightarrow\alpha_{i}$ and $\hat{a}_{i}^{\dag}\leftrightarrow\alpha_{i}^{\ast}$, we obtain
\begin{eqnarray}
\frac{dP}{dt} &=& \left\{-\kappa\left[\frac{\partial}{\partial\alpha_{1}}\alpha_{2}^{\ast}\alpha_{3}+ \frac{\partial}{\partial\alpha_{1}^{\ast}}\alpha_{2}\alpha_{3}^{\ast}+
\frac{\partial}{\partial\alpha_{2}}\alpha_{1}^{\ast}\alpha_{3}+ \frac{\partial}{\partial\alpha_{2}^{\ast}}\alpha_{1}\alpha_{3}^{\ast}
-\frac{\partial}{\partial\alpha_{3}}\alpha_{1}\alpha_{2}-\frac{\partial}{\partial\alpha_{3}^{\ast}}\alpha_{1}^{\ast}\alpha_{2}^{\ast}\right]\right.\nonumber\\
& &\left.
+\frac{\kappa}{2}\left[\left(\frac{\partial^{2}}{\partial\alpha_{1}\partial\alpha_{2}}+\frac{\partial^{2}}{\partial\alpha_{2}\partial\alpha_{1}}\right)\alpha_{3}+\left(\frac{\partial^{2}}{\partial\alpha_{1}^{\ast}\partial\alpha_{2}^{\ast}}+\frac{\partial^{2}}{\partial\alpha_{2}^{\ast}\partial\alpha_{1}^{\ast}}\right)\alpha_{3}^{\ast}
\right]\right\}P(\vec{\alpha}_{i},\vec{\alpha}_{i}^{\ast},t).
\label{eq:FPE}
\end{eqnarray}
We immediately see that the diffusion term of the above Fokker-Planck equation is not positive-definite, so we will use the positive-P representation, which at the cost of doubling the dimensionality of the phase-space, allows us to map the resulting Fokker-Planck equation onto a set of stochastic differential equations~\cite{plusP}. Making the changes $\alpha_{i}^{\ast}\rightarrow\alpha_{i}^{+}$ and noting that $\alpha_{i}^{+}=\alpha_{i}^{\ast}$ in a distributional sense, we find the set of It\^o calculus~\cite{SMCrispin} equations,
\begin{eqnarray}
\frac{d\alpha_{1}}{dt} &=& \kappa\alpha_{2}^{+}\alpha_{3}+\sqrt{\frac{\kappa\alpha_{3}}{2}}\left(\eta_{1}+i\eta_{3}\right),\nonumber\\
\frac{d\alpha_{1}^{+}}{dt} &=& \kappa\alpha_{2}\alpha_{3}^{+}+\sqrt{\frac{\kappa\alpha_{3}^{+}}{2}}\left(\eta_{2}+i\eta_{4}\right),\nonumber\\
\frac{d\alpha_{2}}{dt} &=& \kappa\alpha_{1}^{+}\alpha_{3}+\sqrt{\frac{\kappa\alpha_{3}}{2}}\left(\eta_{1}-i\eta_{3}\right),\nonumber\\
\frac{d\alpha_{2}^{+}}{dt} &=& \kappa\alpha_{1}\alpha_{3}^{+}+\sqrt{\frac{\kappa\alpha_{3}^{+}}{2}}\left(\eta_{2}-i\eta_{4}\right),\nonumber\\
\frac{d\alpha_{3}}{dt} &=& -\kappa\alpha_{1}\alpha_{2},\nonumber\\
\frac{d\alpha_{3}^{+}}{dt} &=& -\kappa\alpha_{1}^{+}\alpha_{2}^{+},
\label{eq:SFGSDE}
\end{eqnarray}
where the Gaussian noise terms have the correlations
\begin{equation}
\overline{\eta_{j}(t)}=0,\:\:\:\overline{\eta_{j}(t)\eta_{k}(t')}=\delta_{jk}\delta(t-t').
\label{eq:noises}
\end{equation}
We note here that the stochastic positive-P representation equations are an exact mapping from the Hamiltonian and allow the calculation of any normally-ordered operator moments through an averaging process over trajectories, such that
\begin{equation}
\overline{(\alpha_{j}^{+})^{m}\alpha_{k}^{n}}\rightarrow\langle  (\hat{a}_{j}^{\dag})^{m}\hat{a}_{k}^{n}\rangle,
\label{eq:PPdef}
\end{equation}
in the limit of a large number of stochastic trajectories, with the bar representing a classical averaging. We also note that this relationship is only valid where there are no divergence problems with the stochastic integration, and that there were none in any of the results presented here.

It is worthwhile examining the dynamics predicted by the stochastic integration of Eq.~\ref{eq:SFGSDE} as this will provide some insight into the full system where we will consider a doubly-pumped triply resonant cavity. We note here that we will not attempt to analytically solve the mean-field equations found by dropping the noise terms in Eq.~\ref{eq:SFGSDE}, as this procedure is of limited use in second harmonic generation~\cite{SHG1,SHG2,SHG3}, which differs in having the two lower frequency modes as identical.
Hence we consider that linearisation of these equations is unlikely to be accurate after some shortish interaction time for either the mean fields or the quantum correlations. We will use a linearisation procedure in section~\ref{sec:cavidade}, where this is much more useful as long as we are careful with the regime of viability.

\begin{figure}[tbhp]
\includegraphics[width=.75\columnwidth]{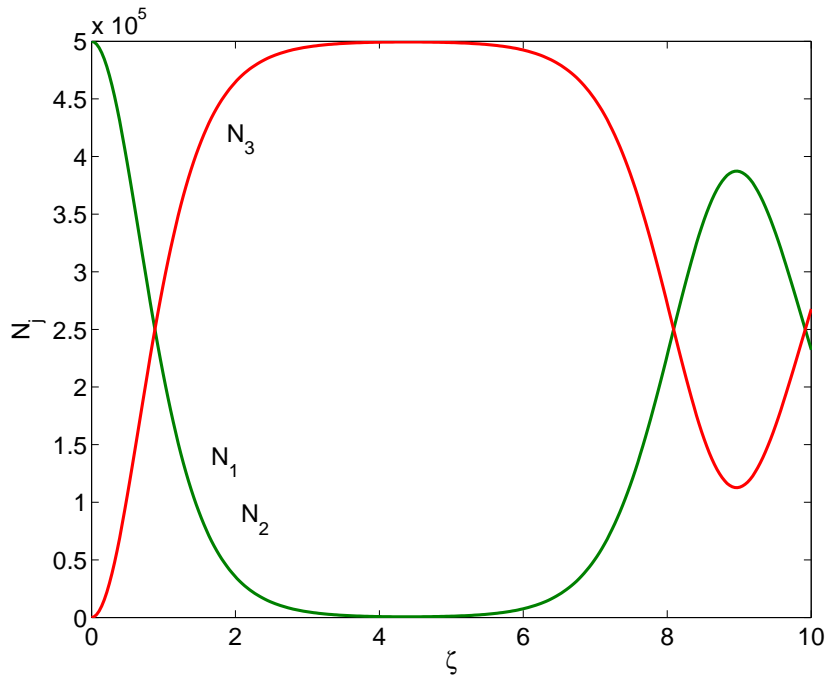}
\caption{(colour online) Mean intensities from an average over $2.67\times 10^{6}$ stochastic trajectories of the positive-P equations (\ref{eq:SFGSDE}). The horizontal axis is a scaled dimensionless time, $\zeta = \kappa|\alpha_{1}(0)|t$. All quantities plotted in this and subsequent graphs are dimensionless.}
\label{fig:ambulante}
\end{figure}

The first quantities we calculate are the mean intensities as a function of interaction time, as shown in Fig.~\ref{fig:ambulante} for $\kappa=0.01$ and the initial conditions $\alpha_{1}(0)=\alpha_{2}(0)=1000/\sqrt{2}$ and $\alpha_{3}(0)=0$. Both the pumping fields are treated as coherent states, which is reasonable for stabilised lasers operating above threshold. We see that the dynamics are reminiscent of those obtained from a similar treatment of second harmonic generation, showing an almost complete conversion to the higher frequency, followed by partial reconversion to the two lower frequencies~\cite{SHG2}. We note here that a previous semiclassical analysis for equal intensities in the two low frequency inputs showed complete conversion and no subsequent reconversion~\cite{Band}, processes which can be explained as due to quantum fluctuations and the discrete nature of the electromagnetic field~\cite{SHG2,Tindle}. Stochastic integration using the phase-space representations automatically includes these effects and allows for different quantum states of the input fields to be simulated, such as thermal~\cite{pendular} and number states~\cite{stojan}. 

\begin{figure}[tbhp]
\includegraphics[width=.75\columnwidth]{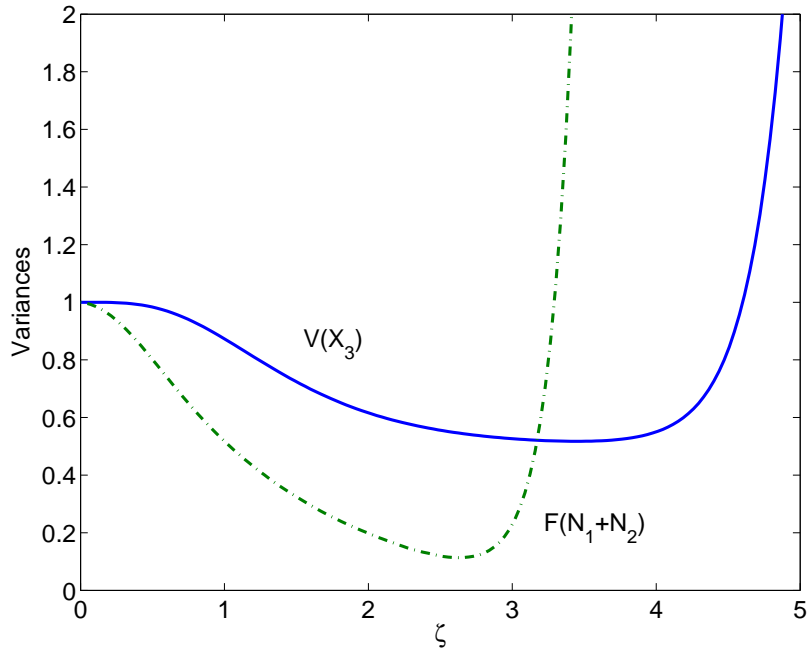}
\caption{(colour online) The $X$ quadrature variance for the field at $\omega_{3}$ and the Fano factor for the intensity sum of the two low frequency fields. A value of less than one signifies squeezing or sub-Poissonian statistics respectively. The two low frequency fields show excessive noise in both $X$ and $Y$ quadratures and exhibit super-Poissonian intensity fluctuations.}
\label{fig:Vprop}
\end{figure}

As our motivation for investigating this system is its utility for producing entangled and other quantum states of electromagnetic fields, we will now look at these quantum properties. The first of these is single-mode squeezing, for which we first need to define quadrature operators. We will define a general quadrature of the field at a given quadrature angle as~\cite{Danbook}
\begin{equation}
\hat{X}_{j}(\theta) = \hat{a}_{j}\e^{-i\theta}+\hat{a}_{j}^{\dag}\e^{i\theta},
\label{eq:quaddef}
\end{equation}
and use the shorthand $\hat{X}_{j}(0)=\hat{X}_{j}$ and $\hat{X}_{j}(\frac{\pi}{2})=\hat{Y}_{j}$. With these definitions, a squeezed state is one for which $V(\hat{X}(\theta))<1$. In Fig.~\ref{fig:Vprop} we show the results of stochastic calculations for the variances of both the high frequency $X_{3}$ quadrature and the sum of the low-frequency intensities. For the latter, which has been predicted to give sub-Poissonian statistics in the intracavity case~\cite{Eschmann}, we have plotted the Fano factor, defined for this case as
\begin{equation}
F(N_{1}+N_{2}) = \frac{V(N_{1}+N_{2})}{N_{1}+N_{2}},
\label{eq:Fano}
\end{equation}
so that any value below one means that the combined mode has less intensity fluctuations than a coherent state. Comparing the two curves in Fig.~\ref{fig:Vprop}, we see that the squeezing in $\hat{X}_{3}$ essentially continues until the upconversion turns into downconversion, whereas the noise suppression in the intensity sum disappears as the low frequency intensities approach their minima. The other quadratures, the single-mode intensities and the intensity differences all exhibit excess noise as the interaction proceeds. This ability to produce quadrature squeezed light at the sum frequency could be useful in, for example, super resolution optical measurements~\cite{Fabre}. 

\begin{figure}[tbhp]
\includegraphics[width=.75\columnwidth]{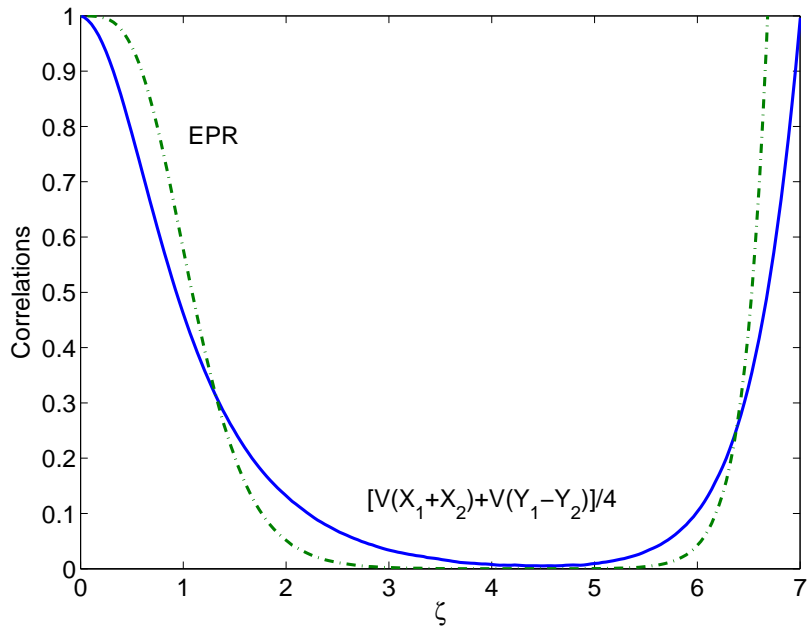}
\caption{(colour online) The Duan-Simon correlation, $V(X_{1}+X_{2})+V(Y_{1}-Y_{2})$, and the Reid EPR correlation, $V_{inf}(X_{j})V_{inf}(Y_{j})$, for the two low frequency fields. With symmetric inputs, the EPR correlation is equal for $j=1$ and $2$. Note that the Duan-Simon value has been divided by four so that both entanglement and the EPR paradox are demonstrated for values less than $1$.}
\label{fig:EPRDuanprop}
\end{figure}

The correlation between the intensities of the two low frequency fields suggests that they may exhibit bipartite entanglement. We examine this using the criteria for continuous variables developed by Duan \etal~\cite{Duan} and Simon~\cite{Simon}. In the present case where the two pump modes have equal intensities, it is sufficient to violate one of the inequalities,
\begin{equation}
V(\hat{X}_{1}\pm\hat{X}_{2})+V(\hat{Y}_{1}\mp\hat{Y}_{2})\leq 4,
\label{eq:DuanSimon}
\end{equation}
to establish that the two low frequency modes are entangled. It is known that entangled states are a superset of states which can be used for steering~\cite{steering} and that this can be demonstrated in the continuous variable case by violation of the Reid EPR (Einstein-Podolsky-Rosen) inequalities~\cite{ReidEPR1,ReidEPR2}. In this case the inequality can be written as
\begin{equation}
V_{inf}(\hat{X}_{j})V_{inf}(\hat{Y}_{j})\geq 1,
\label{eq:MargaretEPR}
\end{equation}
where 
\begin{eqnarray}
V_{inf}(\hat{X}_{j}) &=& V(\hat{X}_{j})-\frac{\left[V(\hat{X}_{j},\hat{X}_{k})\right]^{2}}{V(\hat{X}_{k})},\nonumber\\
V_{inf}(\hat{Y}_{j}) &=& V(\hat{Y}_{j})-\frac{\left[V(\hat{Y}_{j},\hat{Y}_{k})\right]^{2}}{V(\hat{Y}_{k})},
\label{eq:Vinf}
\end{eqnarray}
with violation of the inequality showing that modes $j$ and $k$ are entangled in the EPR sense. 
As can be seen in Fig.~\ref{fig:EPRDuanprop}, where we have divided the correlation for the sum of the $\hat{X}$ quadratures and the difference of the $\hat{Y}$ quadratures by $4$, these two measures give an unambiguous demonstration that the two modes become entangled by the interaction and are in fact entangled in the strong sense required by the concept of steering. Having now established the properties of the Hamiltonian dynamics, we will turn our attention to a more quantitative analysis of the intracavity case. 

\section{Intracavity dynamics}
\label{sec:cavidade}

In the intracavity case we are interested in correlations of the steady-state fields, as these are what are normally measured using homodyne techniques. To analyse these theoretically, we
must add pumping and loss terms to the equations given above~(\ref{eq:SFGSDE}). Making the usual zero temperature Born and Markov approximations~\cite{Danbook} for the interactions with the reservoirs, we find
\begin{eqnarray}
\frac{d\alpha_{1}}{dt} &=& \epsilon_{1}-\gamma_{1}\alpha_{1}+ \kappa\alpha_{2}^{+}\alpha_{3}+\sqrt{\frac{\kappa\alpha_{3}}{2}}\left(\eta_{1}+i\eta_{3}\right),\nonumber\\
\frac{d\alpha_{1}^{+}}{dt} &=& \epsilon_{1}^{\ast}-\gamma_{1}\alpha_{1}^{+}+ \kappa\alpha_{2}\alpha_{3}^{+}+\sqrt{\frac{\kappa\alpha_{3}^{+}}{2}}\left(\eta_{2}+i\eta_{4}\right),\nonumber\\
\frac{d\alpha_{2}}{dt} &=& \epsilon_{2}-\gamma_{2}\alpha_{2}+ \kappa\alpha_{1}^{+}\alpha_{3}+\sqrt{\frac{\kappa\alpha_{3}}{2}}\left(\eta_{1}-i\eta_{3}\right),\nonumber\\
\frac{d\alpha_{2}^{+}}{dt} &=& \epsilon_{2}^{\ast}-\gamma_{2}\alpha_{2}^{+}+ \kappa\alpha_{1}\alpha_{3}^{+}+\sqrt{\frac{\kappa\alpha_{3}^{+}}{2}}\left(\eta_{2}-i\eta_{4}\right),\nonumber\\
\frac{d\alpha_{3}}{dt} &=& -\gamma_{3}\alpha_{3} -\kappa\alpha_{1}\alpha_{2},\nonumber\\
\frac{d\alpha_{3}^{+}}{dt} &=& -\gamma_{3}\alpha_{3}^{+}-\kappa\alpha_{1}^{+}\alpha_{2}^{+},
\label{eq:cavP+}
\end{eqnarray}
where the $\gamma_{j}$ are the cavity loss rates at the respective frequencies, the $\epsilon_{j}$ are coherent pumping terms, and the noise terms have the same correlations as in Eq.~\ref{eq:SFGSDE}.

In the intracavity case it is often possible to find the appropriate fluctuation spectra via a process of linearising the fluctuations in the variables about their steady-state, classical solutions~\cite{Danbook}. This enables us to write equations for the fluctuations as a multivariate Ornstein-Uhlenbeck process, from which it is particularly simple to extract the appropriate spectral correlations. There are however, two caveats which must be considered here. The first is that the steady-state solutions of the classical equations (Eq.~\ref{eq:cavP+} with the noise terms dropped) must be the actual mean-field solutions, and the second is that the drift matrix in the resulting equation for the Ornstein-Uhlenbeck process must not have any negative real parts to its eigenvalues. These conditions are known to occur with intracavity parametric processes, with the first being violated in the self-pulsing regime of second harmonic generation~\cite{Haken,McNeil,mepulse} and the second being a problem at the critical operating point of the optical parametric oscillator~\cite{mjcnonlinear}. In order to check the stability we first turn to solving the steady-state equations for the mean fields.

Although in general not possible, we can find analytical solutions for the steady states in the case where we set the pumping strengths equal, with $\epsilon_{1}=\epsilon_{2}=\epsilon$, and the two low frequency loss rates also equal, with $\gamma_{1}=\gamma_{2}=\gamma$. Although achieving equal loss rates may not be so simple in the laboratory, it is useful here in order to gain some insight. With these values set equal, the two low frequency modes will have equal amplitudes with $\alpha_{1}^{ss}=\alpha_{2}^{ss}=\alpha^{ss}$, which leads to a cubic equation for $\alpha_{3}^{ss}$ (note that we will now drop the superscripts for notational convenience),
\begin{equation}
\gamma_{3}\kappa^{2}\alpha_{3}^{2}-2\gamma\gamma_{3}\kappa\alpha_{3}^{2}+\gamma_{3}\gamma^{2}\alpha_{3}+\kappa\epsilon^{2}=0,
\label{eq:ssalpha3}
\end{equation}
along with a simple expression for $\alpha$,
\begin{equation}
\alpha = \frac{\epsilon}{\gamma-\kappa\alpha_{3}}.
\label{eq:ssalpha}
\end{equation}
We note that there is no divergence here in the value of $\alpha$ since the high frequency field amplitude is negative in the steady state.
Without setting these loss and pumping rates equal, it is much more difficult to find analytical solutions, although we can proceed via numerics, which we will do later in order to examine the effects of asymmetry. In the present case where the cavity is triply resonant, $\alpha_{3}$ will be the real solution of Eq.~\ref{eq:ssalpha3}. This is found as
\begin{equation}
\alpha_{3} = \frac{1}{6}\left[\frac{4\gamma}{\kappa}+\frac{16^{1/3}\gamma^{2}\gamma_{3}}{\xi}+\frac{2^{2/3}\xi}{\gamma_{3}\kappa^{2}}\right],
\label{eq:cubicsol}
\end{equation}
where
\begin{equation}
\xi = \left[-2\gamma^{3}\gamma_{3}^{3}\kappa^{3}-27\gamma_{3}^{2}\kappa^{5}\epsilon^{2}+\sqrt{27\gamma_{3}^{4}\kappa^{8}\epsilon^{2}\left(27\kappa^{2}\epsilon^{2}+4\gamma_{3}\gamma^{3}\right)}\right]^{1/3}.
\label{eq:horriblething}
\end{equation}

We now decompose the variables into their steady-state classical values and fluctuations around these,
\begin{equation}
\alpha_{j}=\alpha_{j}^{ss}+\delta\alpha_{j},
\label{eq:decomposition}
\end{equation}
and find the equations of motion for the fluctuation vector,
\begin{equation}
\delta X = \left[\delta\alpha_{1},\delta\alpha_{1}^{+},\delta\alpha_{1},\delta\alpha_{2}^{+},\delta\alpha_{3},\delta\alpha_{3}^{+}\right]^{T},
\label{eq:flukvek}
\end{equation}
as 
\begin{equation}
\delta X = -Adt+BdW,
\label{eq:ornsteinandhismate}
\end{equation}
where $A$ is the drift matrix (remembering that the $\alpha_{j}$ are now to be read as the steady-state values),
\begin{equation}
A=\left[\begin{array}{cccccc}
\gamma_{1} & 0 & 0 & -\kappa\alpha_{3} & -\kappa\alpha_{2}^{\ast} & 0\\
0 & \gamma_{1} & -\kappa\alpha_{3}^{\ast} & 0 & 0 & -\kappa\alpha_{2}\\
0 & -\kappa\alpha_{3} & \gamma_{2} & 0 & -\kappa\alpha_{1}^{\ast} & 0\\
-\kappa\alpha_{3}^{\ast} & 0 &0 & \gamma_{2} & 0 & -\kappa\alpha_{1}\\
\kappa\alpha_{2} & 0 & \kappa\alpha_{1} & 0 & \gamma_{3} & 0\\
0 & \kappa\alpha_{2}^{\ast} & 0 & \kappa\alpha_{1}^{\ast} & 0 & \gamma_{3}\end{array}\right],
\label{eq:Amat}
\end{equation}
$dW$ is a vector of real Wiener increments and $B$ is the diffusion matrix,
\begin{equation}
B=\left[\begin{array}{cccccc}
\sqrt{\frac{\kappa\alpha_{3}}{2}} & 0 & i\sqrt{\frac{\kappa\alpha_{3}}{2}} & 0 & 0\\
0 & \sqrt{\frac{\kappa\alpha_{3}^{\ast}}{2}} & 0 & i\sqrt{\frac{\kappa\alpha_{3}^{\ast}}{2}} & 0 & 0\\
\sqrt{\frac{\kappa\alpha_{3}}{2}} & 0 & -i\sqrt{\frac{\kappa\alpha_{3}}{2}} & 0 & 0\\
0 & \sqrt{\frac{\kappa\alpha_{3}^{\ast}}{2}} & 0 & -i\sqrt{\frac{\kappa\alpha_{3}^{\ast}}{2}} & 0 & 0\\
0 & 0 & 0 & 0 & 0 & 0\\
0 & 0 & 0 & 0 & 0 & 0
\end{array}\right].
\label{eq:Bmat}
\end{equation}
In parameter regimes where the matrix $A$ has no negative real part to any of its eigenvalues, we may simply find the intracavity spectra via the relation
\begin{equation}
S(\omega)=\left(A+i\omega\openone\right)^{-1}BB^{\text{T}}\left(A^{\text{T}}-i\omega\openone\right)^{-1},
\label{eq:inspek}
\end{equation}
after which we use the standard input-output relations~\cite{mjc}
to relate these to the experimentally measurable quantities outside the cavity.

\begin{figure}[tbhp]
\includegraphics[width=.75\columnwidth]{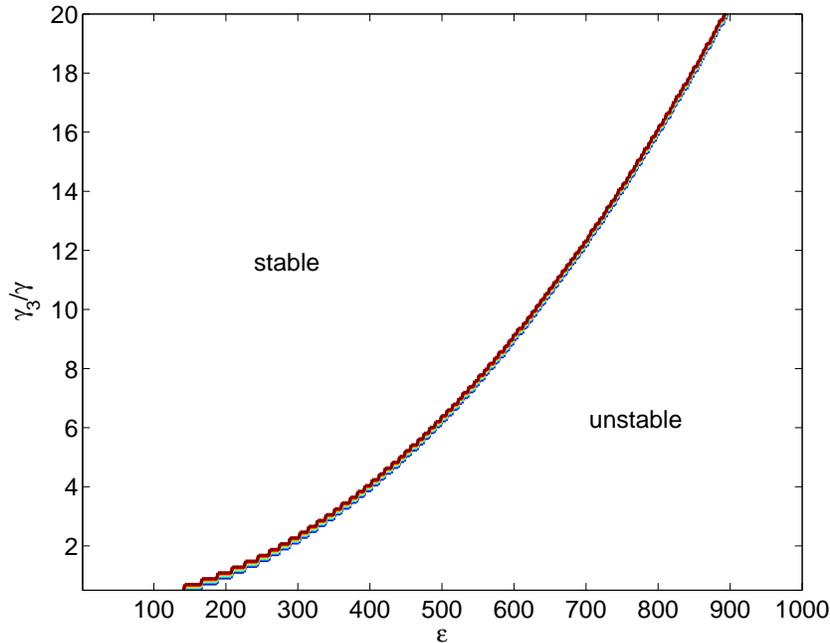}
\caption{(colour online) The regions of stability and instability as a function of $\epsilon$ and $\gamma_{3}/\gamma$ for $\kappa=10^{-2}$ and $\gamma=\gamma_{1}=\gamma_{2}=1$. In the region to the right of the line there is always at least one eignevalue with a negative real part and the linearisation process for calculating spectra is not valid.}
\label{fig:estabilidade}
\end{figure}

Again setting $\epsilon_{1}=\epsilon_{2}=\epsilon$, and the two low frequency loss rates also equal, with $\gamma_{1}=\gamma_{2}=\gamma$, so that $\alpha_{1}=\alpha_{2}=\alpha$ and all the intracavity fields are real, we find analytical solutions for the eigenvalues as
\begin{eqnarray}
\lambda_{1,2} &=& \gamma\pm\kappa\alpha_{3},\nonumber\\
\lambda_{3,4} &=& \frac{1}{2}\left[\gamma+\gamma_{3}+\kappa\alpha_{3}\pm\sqrt{(\gamma-\gamma_{3})^{2}+2\kappa\alpha_{3}(\gamma-\gamma_{3})+\kappa^{2}(\alpha_{3}^{2}-8\alpha^{2})}\right],\nonumber\\
\lambda_{5,6} &=& \frac{1}{2}\left[\gamma+\gamma_{3}-\kappa\alpha_{3}\pm\sqrt{(\gamma-\gamma_{3})^{2}-2\kappa\alpha_{3}(\gamma-\gamma_{3})+\kappa^{2}(\alpha_{3}^{2}-8\alpha^{2})}\right].
\label{eq:autovalores}
\end{eqnarray}
Since $\alpha_{3}$ is negative in the steady-state, there is an obvious instability when $\alpha_{3}<-\gamma/\kappa$, found from the first pair of eigenvalues. Unlike the case of the optical parametric oscillator, where a simple expression can be found for the critical pumping value, we find that it is easiest here to express this in terms of the low frequency field amplitudes, defining a critical field,
\begin{equation}
\alpha_{c} = \frac{\epsilon}{2\gamma}.
\label{eq:alphacrit}
\end{equation}
We note here that this simple expression will not hold unless we set the pumping and loss rates symmetrically, as we did for our analytic eigenvalue analysis. Fig.~\ref{fig:estabilidade} shows how the parameter space is divided into stable and unstable regions in this symmetric case, as we vary the pumping and the ratio between the high and low frequency loss rates. All the spectral results we give will be from the stable region. 
We will return to the region above $\alpha_{c}$ below, using stochastic integration to solve the full equations of motion. In the meantime we will examine the quantum correlations in the stable region of the parameter space.

\section{Entanglement properties}
\label{eq:emaranhado}

\begin{figure}[tbhp]
\includegraphics[width=.75\columnwidth]{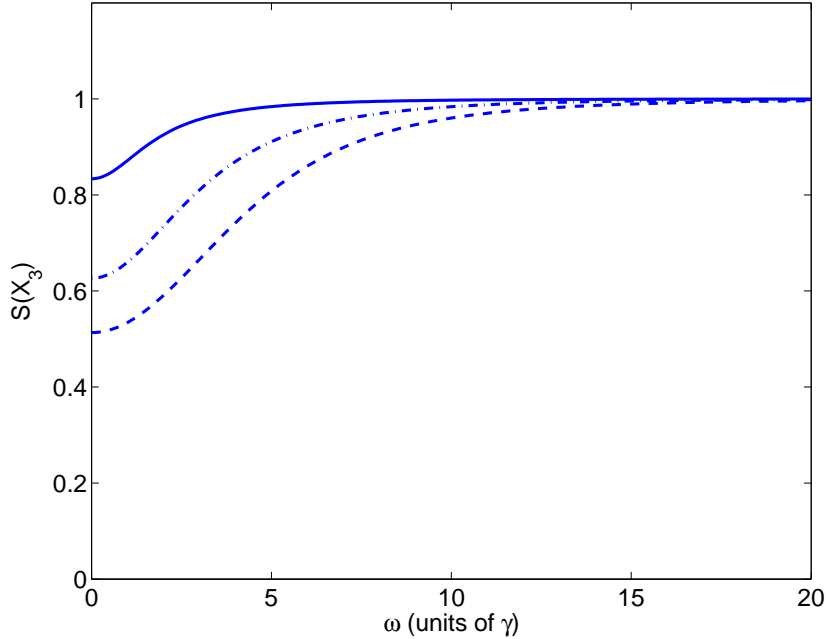}
\caption{(colour online) The spectral variance of the high frequency quadrature, $\hat{X}_{3}$, for $\kappa=10^{-2}$, $\gamma_{1}=\gamma_{2}=\gamma=1$, and $\gamma_{3}=10$. The solid line is for $\epsilon_{1}=\epsilon_{2}=200$, giving $\alpha_{1}=\alpha_{2}=0.25\alpha_{c}$, the dash-dotted line is for $\epsilon_{1}=\epsilon_{2}=400$, giving $\alpha_{1}=\alpha_{2}=0.61\alpha_{c}$, and the dashed line is for $\epsilon_{1}=\epsilon_{2}=600$, giving $\alpha_{1}=\alpha_{2}=0.95\alpha_{c}$.}
\label{fig:VX3cav}
\end{figure}

\begin{figure}[tbhp]
\includegraphics[width=.75\columnwidth]{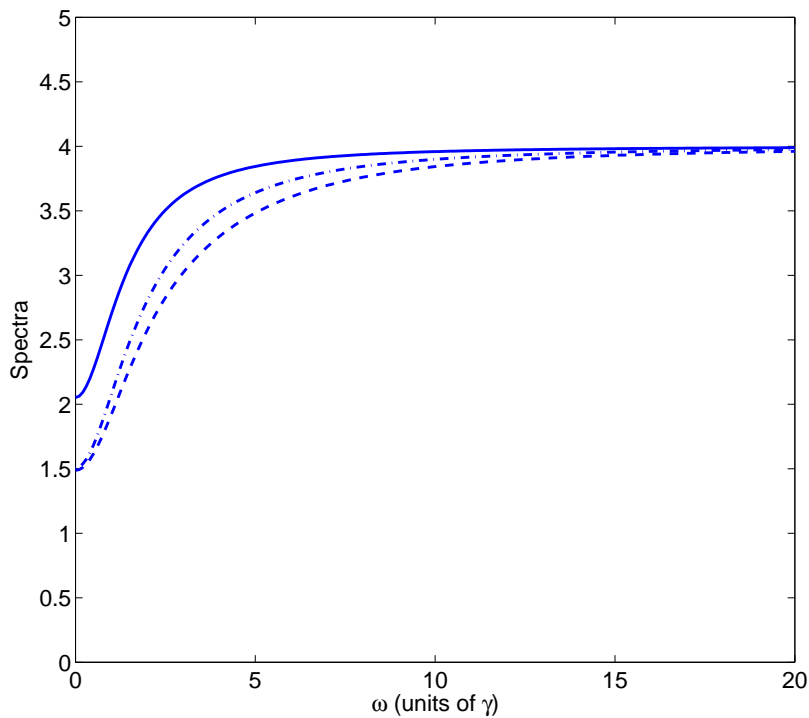}
\caption{(colour online) The Duan-Simon entanglement correlations for the low frequency modes $1$ and $2$ and the same parameters as Fig.~\ref{fig:VX3cav}. The two quadratures which violate the inequality (less than $4$) and hence signify entanglement are $X_{1}+X_{2}$ and $Y_{1}-Y_{2}$. The solid line is for $\epsilon_{1}=\epsilon_{2}=200$, giving $\alpha_{1}=\alpha_{2}=0.25\alpha_{c}$, the dash-dotted line is for $\epsilon_{1}=\epsilon_{2}=400$, giving $\alpha_{1}=\alpha_{2}=0.61\alpha_{c}$, and the dashed line is for $\epsilon_{1}=\epsilon_{2}=600$, giving $\alpha_{1}=\alpha_{2}=0.95\alpha_{c}$.}
\label{fig:Duancav}
\end{figure}

In order to examine the entanglement and squeezing properties of this system, we must go beyond the intensity fluctuations calculated previously~\cite{Eschmann} and calculate the phase dependent quadrature correlations given above in Eq.~\ref{eq:DuanSimon} and Eq.~\ref{eq:MargaretEPR}. The first of these which we show, in Fig.~\ref{fig:VX3cav}, is the output spectral variance of the high frequency amplitude quadrature, $\hat{X}_{3}$. We see that, as in the travelling wave case, this is squeezed and that the degree of squeezing increases as we approach $\alpha_{c}$. As shown in Fig.~\ref{fig:Duancav}, this system is also a good source of bichromatic bipartite entanglement between the two low frequency fields. We see that the degree of violation of the Duan-Simon inequality also increases as we approach $\alpha_{c}$, although not as markedly as does the high frequency squeezing. Consistent with the fact that SFG suppresses the noise in the sum of the two low frequency intensities, unlike parametric downconversion which suppresses the noise in the difference, we find that the entanglement is signalled by the correlation $V(\hat{X}_{1}+\hat{X}_{2})+V(\hat{Y}_{1}-\hat{Y}_{2})$. We did not find any evidence of tripartite entanglement in this system, at least for the parameter ranges we have investigated. 

\subsection{Symmetric and asymmetric bipartite steering}
\label{subsec:steering}

In a recent article~\cite{steering}, Wiseman \etal~ discussed the concept of steering, which had originally been introduced by Schr\"odinger~\cite{schrodinger1,schrodinger2} in the context of the EPR paradox~\cite{albert}. Wiseman \etal~ explain this concept in terms of a bipartite state prepared by Alice, who then sends one part to Bob, with this process being repeatable. After measuring their own parts, they communicate classically, with Alice trying to convince Bob that the prepared state is entangled. If Bob cannot explain the correlations using any local hidden state (LHS) model, the state must be entangled. Schr\"odinger used the concept of LHS to say that Bob's system could have a definite state before measurement, even though this actual state would be unknown to Bob. He introduced the concept of steering to describe how Alice could affect Bob's state via her choice of measurement basis, but expected that this would never be seen experimentally. 

\begin{figure}[tbhp]
\includegraphics[width=.75\columnwidth]{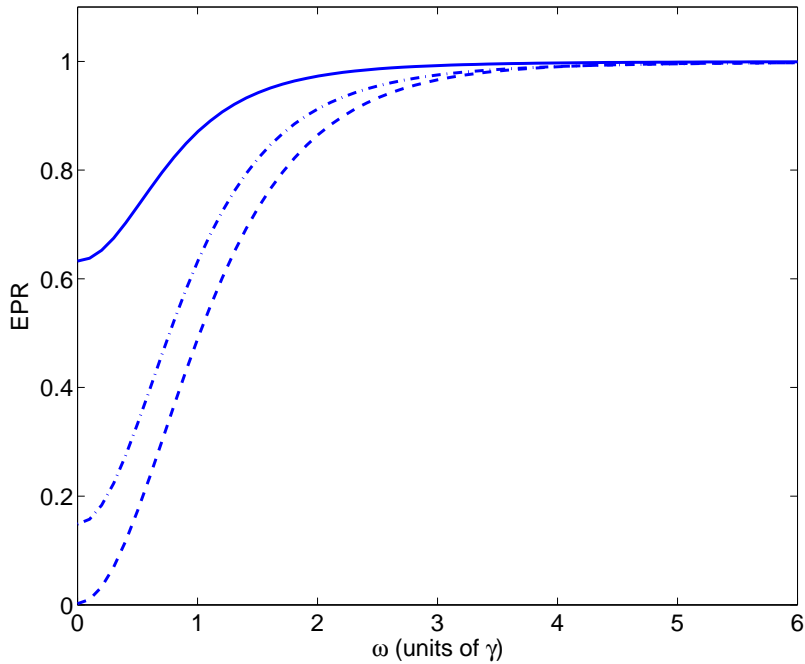}
\caption{(colour online) The spectral EPR product of the inferred variances for modes $1$ and $2$, and the same parameters as Fig.~\ref{fig:VX3cav}. The solid line is for $\epsilon_{1}=\epsilon_{2}=200$, giving $\alpha_{1}=\alpha_{2}=0.25\alpha_{c}$, the dash-dotted line is for $\epsilon_{1}=\epsilon_{2}=400$, giving $\alpha_{1}=\alpha_{2}=0.61\alpha_{c}$, and the dashed line is for $\epsilon_{1}=\epsilon_{2}=600$, giving $\alpha_{1}=\alpha_{2}=0.95\alpha_{c}$.}
\label{fig:EPRcav}
\end{figure}

\begin{figure}[tbhp]
\includegraphics[width=.75\columnwidth]{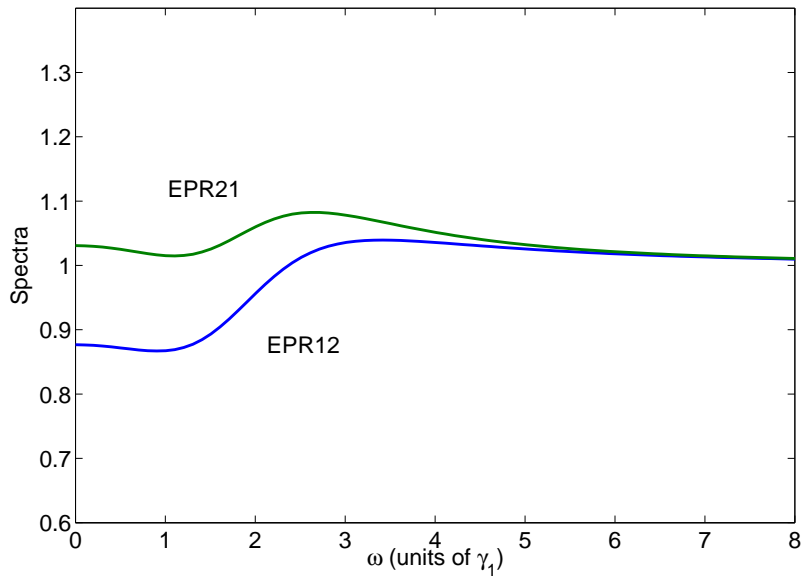}
\caption{(colour online) Demonstration of asymmetric steering for the parameters $\gamma_{1}=1,\:\gamma_{2}=40,\:\gamma_{3}=2,\:\epsilon_{1}=400,\:\epsilon_{2}=2400$ and $\kappa=0.01$. The correlation EPR12 shows that $2$ can steer $1$, while EPR21 shows that $1$ cannot steer $2$.}
\label{fig:womandriver}
\end{figure}

For the purposes of this paper, it is interesting to note that a demonstration of the EPR paradox using the Reid criteria~\cite{ReidEPR1}, as was first done by Ou \etal~\cite{Ou} using parametric downconversion, is equivalent to a demonstration of steering. In Fig.~\ref{fig:EPRcav} we show that this system does allow for demonstrations of steering, with a degree of violation of the Reid inequality (\ref{eq:MargaretEPR}) that increases as the pumping is increased so that $\alpha_{c}$ is approached. For the symmetric inputs and parameters we have chosen here, the steering is totally symmetric, with either Alice ($\omega_{1}$) being able to steer Bob ($\omega_{2}$) or Bob being able to steer Alice. However, in the work of Wiseman \etal, the reader is left with an open question as to whether there exist asymmetric states that are steerable by Alice but not by Bob~\cite{steering}. It seems intuitively obvious that these type of states would not be produced in any normal downconversion processes, which are inherently symmetric in their production of correlated pairs of photons. The freedom we have with the present system, where the pumping rates and mirror losses can be different at the two low frequencies, leads us to expect that it should be a good candidate for the exhibition of asymmetric steering. Indeed, as shown in Fig.~\ref{fig:womandriver}, it is in principle possible to arrange the cavity loss rates and pumpings so that this is seen. In the figure, we have used the labels EPRij to represent the product $V_{inf}(\hat{X}_{i})V_{inf}(\hat{Y}_{i})$ and have chosen a regime where the linearisation process is valid. We also note here that previous coupled systems which have been predicted to produce entangled outputs via evanescently coupled intracavity $\chi^{(2)}$ and $\chi^{(3)}$ processes could also be arranged asymmetrically and thus may also be good candidates for the production of asymmetric steering~\cite{bache,couple1,couple2,nlc,ReidEPR2}. Having shown that asymmetric steering is possible, we may pose another open question as to what it may prove useful for, with one way quantum cryptography being one suggestion, although further development of this theme is outside the scope of this paper. 

\subsection{Quantum dynamics}
\label{subsec:quantumdynamics}

\begin{figure}[tbhp]
\includegraphics[width=.75\columnwidth]{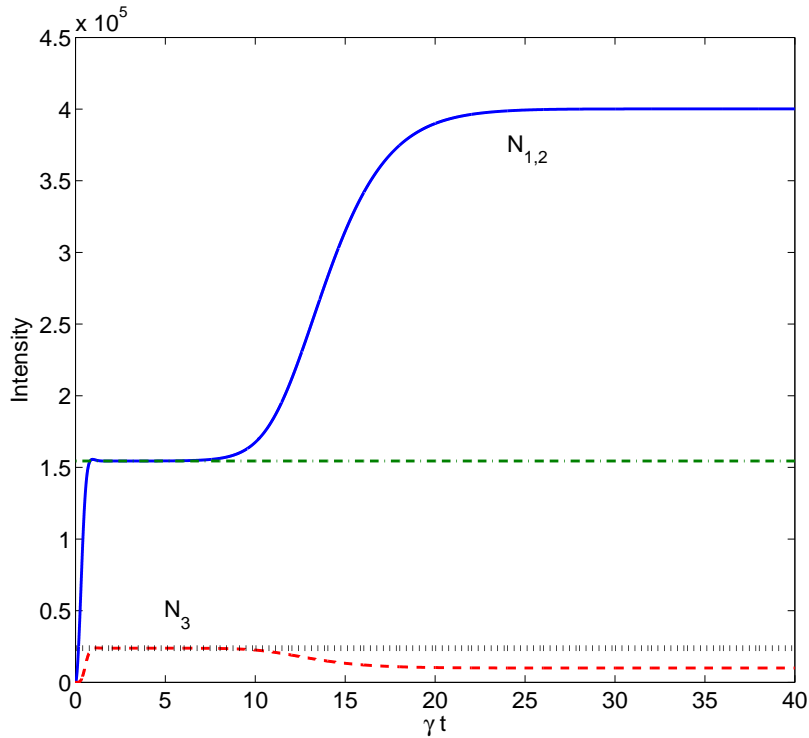}
\caption{(colour online) Intracavity intensities predicted by averaging $2.2\times 10^{5}$ stochastic trajectories of the full positive-P representation equations, for $\gamma_{1}=\gamma_{2}=1,\:\gamma_{3}=10,\:\epsilon_{1}=\epsilon_{2}=10^{3}$ and $\kappa=0.01$. The straight lines are the semiclassical steady-state predictions for the same parameters.}
\label{fig:diferente}
\end{figure}

In the unstable parameter regime as shown in Fig.~\ref{fig:estabilidade} we must resort to stochastic integration of the full equations (\ref{eq:cavP+}), without linearising about the semiclassical solutions. What we find is qualitatively different from the behaviour in the stable regime, as can be seen in Fig.~\ref{fig:diferente}. For pumping strengths for which the semiclassical solution has $\alpha<\alpha_{c}$, the full quantum solutions converge to the semiclassical predictions, as they do in the transient regime of Fig.~\ref{fig:diferente}. However, when the cavity is pumped more strongly, the fields change radically after a short time, with the low frequency fields increasing and the high frequency intensity decreasing as the process of downconversion back into the two low frequencies becomes dominant. While interesting from the point of view of being a system where the semiclassical solutions do not capture the dynamics of the mean fields, we found no evidence of steady-state entanglement or quadrature squeezing in our investigations of this regime. 

\section{Conclusion}

In conclusion we have shown that the system of intracavity sum frequency generation is potentially a versatile and tuneable source of squeezed light, bright bichromatic continuous variable bipartite entanglement, and two mode Einstein-Podolsky-Rosen states.
With recent advances in the fabrication of nonlinear crystals and with the techniques necessary for what we may term its classical uses very well established, this process may become a powerful tool in the optical arsenal of quantum information. The potential ability to run the system in a highly asymmetric manner means that it could be used as an instrument for research into the fundamentals of quantum mechanics in ways that are not possible with the familiar optical parametric oscillator. The ability to choose and combine a range of frequencies, with the combination frequency being quadrature squeezed, may prove useful in super resolution measurements.

\section*{Acknowledgments}

This work was supported by the Australian Research Council.


\end{document}